\documentclass[a4paper,10pt]{article}
\pdfoutput=1
\usepackage{graphicx}
\usepackage{fancyheadings}
\usepackage{wrapfig}
\usepackage{epstopdf}
\usepackage[labelfont=bf]{caption}
\usepackage{amssymb,amsmath}
\usepackage[font=small,skip=0pt]{caption}
\usepackage{titling}
\usepackage{color}             
\usepackage{setspace}          
\usepackage{enumitem}          
\usepackage{footmisc} 
\usepackage{subfig}
\usepackage{float}
\usepackage[colon,authoryear]{natbib}
\bibpunct{[}{]}{,}{n}{,}{,}
\begin{document}
\noindent
{\it 11th Hel.A.S Conference}\\
\noindent
{\it Athens, 8-12 September, 2013}\\
\noindent
%
%
CONTRIBUTED POSTER\\
\noindent
\underline{~~~~~~~~~~~~~~~~~~~~~~~~~~~~~~~~
~~~~~~~~}
\vskip .4cm
%
%
\begin{center}
{\Large\bf
On the dynamics of the three dimensional planetary systems}
\vskip 0.3cm
%
%
{\it
Kyriaki I. Antoniadou, George Voyatzis, John D. Hadjidemetriou$^\dagger$}\\
\texttt{kyant@auth.gr, voyatzis@auth.gr}\\
%
%
Section of Astrophysics, Astronomy and Mechanics, Department of Physics, \\Aristotle University of Thessaloniki, 54124, Greece\\
\end{center}
\vskip 0.3cm\vspace{-.5em}
{\bf Abstract:}
Over the last decades, there has been a tremendous increase in research on extrasolar planets. Many exosolar systems, which consist of a Star and two inclined Planets, seem to be locked in 4/3, 3/2, 2/1, 5/2, 3/1 and 4/1 mean motion resonance (MMR). We herewith present the model used to simulate three dimensional planetary systems and provide planar families of periodic orbits (PO), which belong to all possible configurations that each MMR has, along with their linear horizontal and vertical stability. We focus on depicting stable spatial families (most of them up to mutual inclination of $60^\circ$) generated by PO of planar circular families, because the trapping in MMR could be a consequence of planetary migration process. We attempt to connect the linear stability of PO with long-term stability of a planetary system close to them. This can stimulate the search of real planetary systems in the vicinity of stable spatial PO-counterbalanced by the planets' orbital elements, masses and MMR; all of which could constitute a suitable environment convenient to host them.

\vspace{-.5em}
\section{Model}
\begin{wrapfigure}[23]{L}{3.8cm}\vspace{-1.5em}
\includegraphics[width=3.7cm]{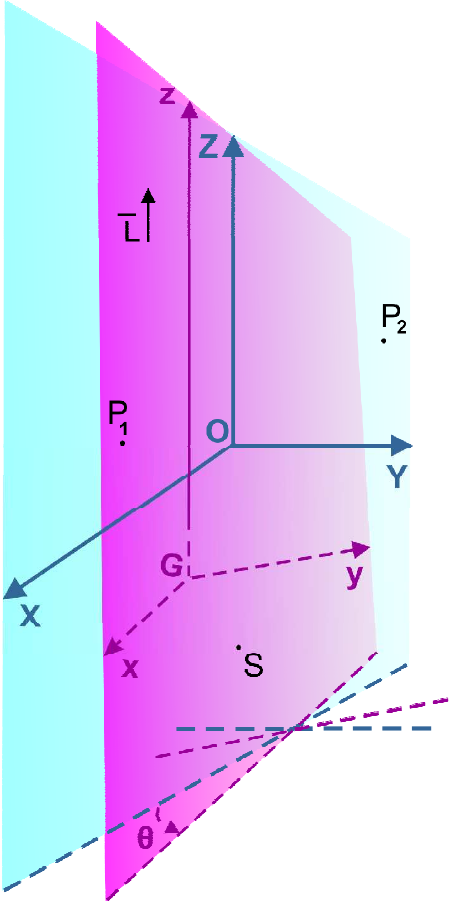}
\caption{Inertial and rotating frame of reference.}
  \label{model}
   \end{wrapfigure}
We introduce a three dimensional system that consists of a Star, $S$ and two inclined planets, $P_1$ and $P_2$, of masses $m_0$, $m_1$ and $m_2$, respectively, which are considered as point masses. The three bodies move in space {\color{cyan}$OXYZ$} ({\color{cyan}inertial frame}) under their mutual gravitational attraction, where the origin $O$ is their fixed center of mass and its $Z$-axis is parallel to the constant angular momentum vector, ${\bf L}$, of the system. The system is described by six degrees of freedom, which can be reduced to four by introducing a suitable {\color{magenta}rotating frame} of reference, {\color{magenta}$Gxyz$}, (Fig. \ref{model}) \citep{av12,av13}, such that:
\begin{enumerate} [noitemsep,nolistsep]
	\item Its origin coincides with the center of mass $G$ of the bodies $S$ and $P_1$.
	\item Its $z$-axis is always parallel to the $Z$-axis.
	\item $S$ and $P_1$ move always on $xz$-plane.
\end{enumerate}
The Lagrangian of the system in the rotating frame of reference is:\vspace{-0.5em}
\begin{equation}
\begin{array}{l}
\mathfrak{L}=\frac{\displaystyle 1}{\displaystyle 2} \mu[a(\dot x_1^2+\dot z_1^2+x_1^2\dot \theta^2)+\displaystyle b [(\dot x_2^2+\dot y_2^2+\dot z_2^2)+\dot\theta^2(x_2^2+y_2^2)+2\dot\theta(x_2\dot y_2-\dot x_2y_2)]]-V,\nonumber \vspace{-0.5em}
\label{Lagrangian}
\end{array}
\end{equation} 
where \vspace{-0.5em}
\begin{equation}
\begin{array}{l}
V=-\frac{\displaystyle m_0 m_1}{\displaystyle r_{01}}-\frac{\displaystyle m_0 m_2}{\displaystyle r_{02}}-\frac{\displaystyle m_1 m_2}{\displaystyle r_{12}},\;
a=m_1/m_0,\;b=m_2/m,\;\mu=m_0 + m_1,\nonumber \vspace{-0.5em}
\label{V}
\end{array}
\end{equation}
and\vspace{-0.5em}
\begin{equation}
\begin{array}{l}
\displaystyle r_{01}^2=(\displaystyle 1+\displaystyle a)^2(\displaystyle x_1^2+z_1^2),\;
\displaystyle r_{02}^2=(\displaystyle a x_1 +\displaystyle x_2)^2+y_2^2+(\displaystyle a  z_1+z_2)^2,\\
\displaystyle r_{12}^2=(\displaystyle x_1 -\displaystyle x_2)^2+y_2^2+ (z_1-z_2)^2.\nonumber \vspace{-0.5em}
\label{Lr}
\end{array}
\end{equation} \vspace{-.5em}
\vspace{-.5em} 
\section{Continuation of periodic orbits}\vspace{-.5em} 
Having defined the Poincar\'e section plane $\Pi=\{y_2=0,\dot y_2>0\}$ in $Gxyz$, the periodic orbits are the fixed or periodic points of this plane and they satisfy the conditions $\textbf{q}(T)=\textbf{q}_0$, where $T$ is the period$, \textbf{q}=\left\{x_1,x_2,z_2,\dot x_1,\dot x_2,\dot y_2, \dot z_2\right\}$ and $\textbf{q}_0=\textbf{q}(0)$. We consider the symmetries with respect to the $xz$-plane and the $x$-axis \citep{mich} and as a result the initial conditions of a $xz$-symmetric periodic orbit are  \vspace{-1.5em} 
\begin{equation}
\begin{array}{llll} 
\{x_{10},x_{20},z_{20},\dot{y}_{20}\} \:\:\text{and}\:\: \dot x_{10}=\dot x_{20}=\dot z_{20}=0. \nonumber\vspace{-1.5em} 
\label{xzsym}
\end{array}
\end{equation}
and the initial conditions of a $x$-symmetric periodic orbit are  \vspace{-1.5em} 
\begin{equation}
\begin{array}{llll} 
\{x_{10},x_{20},\dot{y}_{20},\dot{z}_{20}\} \:\:\text{and}\:\: \dot x_{10}=\dot x_{20}=z_{20}=0.\nonumber \vspace{-1.5em} 
\label{xsym}
\end{array}
\end{equation}

If $\Delta(T)=\{\xi_{ij}\}$, $i,j=1,2$, is the monodromy matrix of \vspace{-.5em} 
\begin{equation} 
\dot \zeta_1=\zeta_2,\quad \dot \zeta_2=A \zeta_1 +B \zeta_2 \nonumber \vspace{-.5em} 
\label{z22}
\end{equation}
where
$A=-\frac{m m_0}{\mu}[(1-\gamma) d^{-3}_{02}+(a+\gamma)d_{12}^{-3}],\; B=-\frac{m_0 m_2 y_2}{\mu x_1 \dot\theta}(d^{-3}_{02}-d_{12}^{-3}),\; d^2_{12}=(x_1-x_2)^2+y_2^2,\; d^2_{02}=(a x_1+x_2)^2+y_2^2$ and $\gamma=b\frac{x_2 \dot\theta + \dot y_2}{x_1 \dot\theta}$, we can define the {\em vertical stability index} \citep{hen} of a planar periodic orbit of period $T$, as\vspace{-1em} 
\begin{equation}
a_v=\frac{1}{2}(\xi_{11}+\xi_{22}) \nonumber  \vspace{-.5em} 
\label{av2}
\end{equation}
If $|a_v|<1$ or $|a_v|>1$ the orbit is vertical stable or unstable, respectively. 

\section{Results}
\begin{figure}[H]
\begin{center}
$\begin{array}{@{\hspace{-.5em}}ccc}
\includegraphics[width=5.5cm,height=5.5cm]{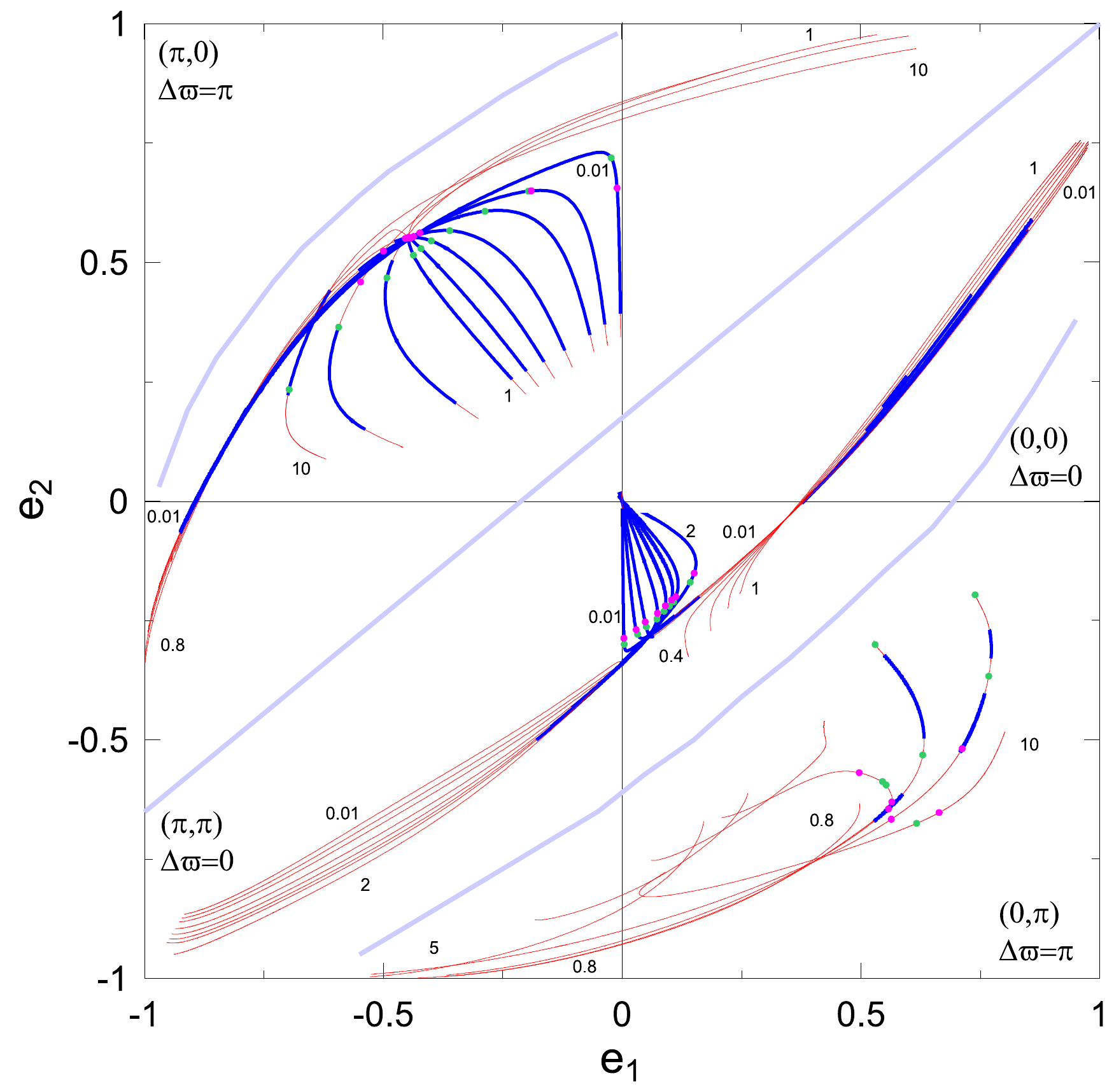}&
\includegraphics[width=5.5cm,height=5.5cm]{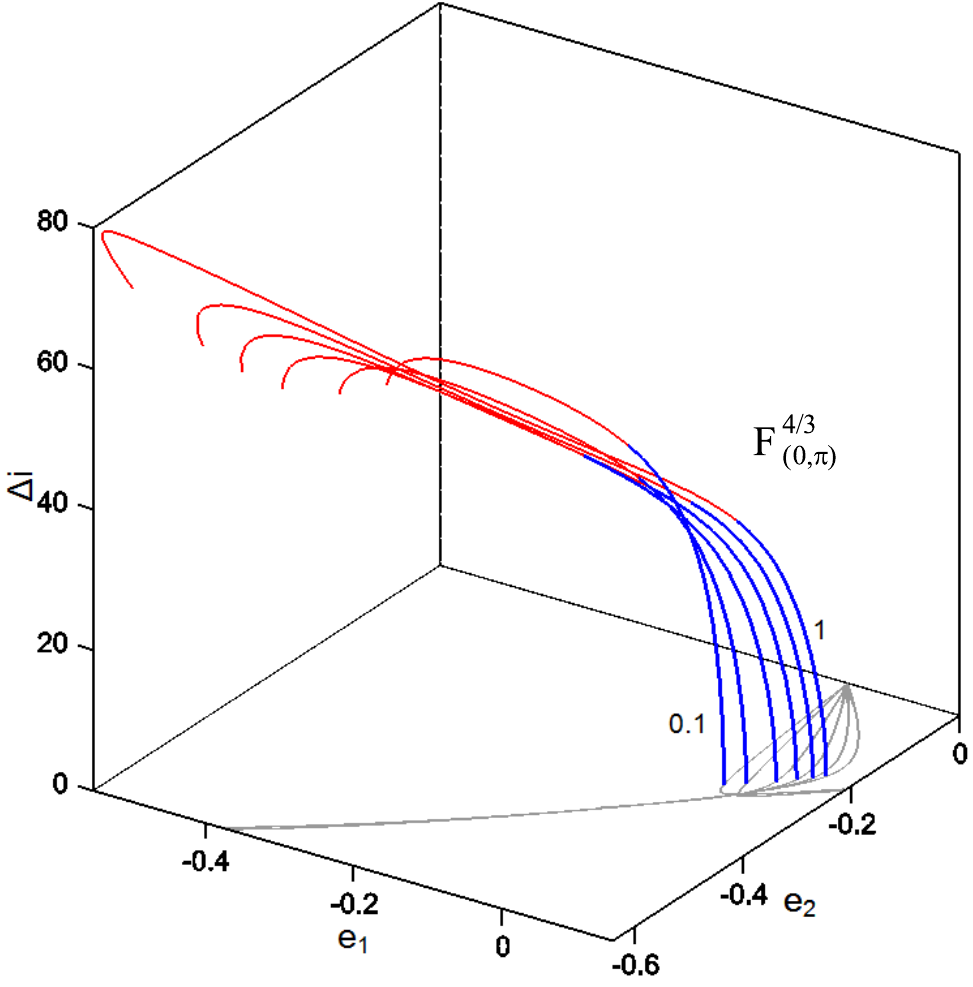}  &
\includegraphics[width=5.5cm,height=5.5cm]{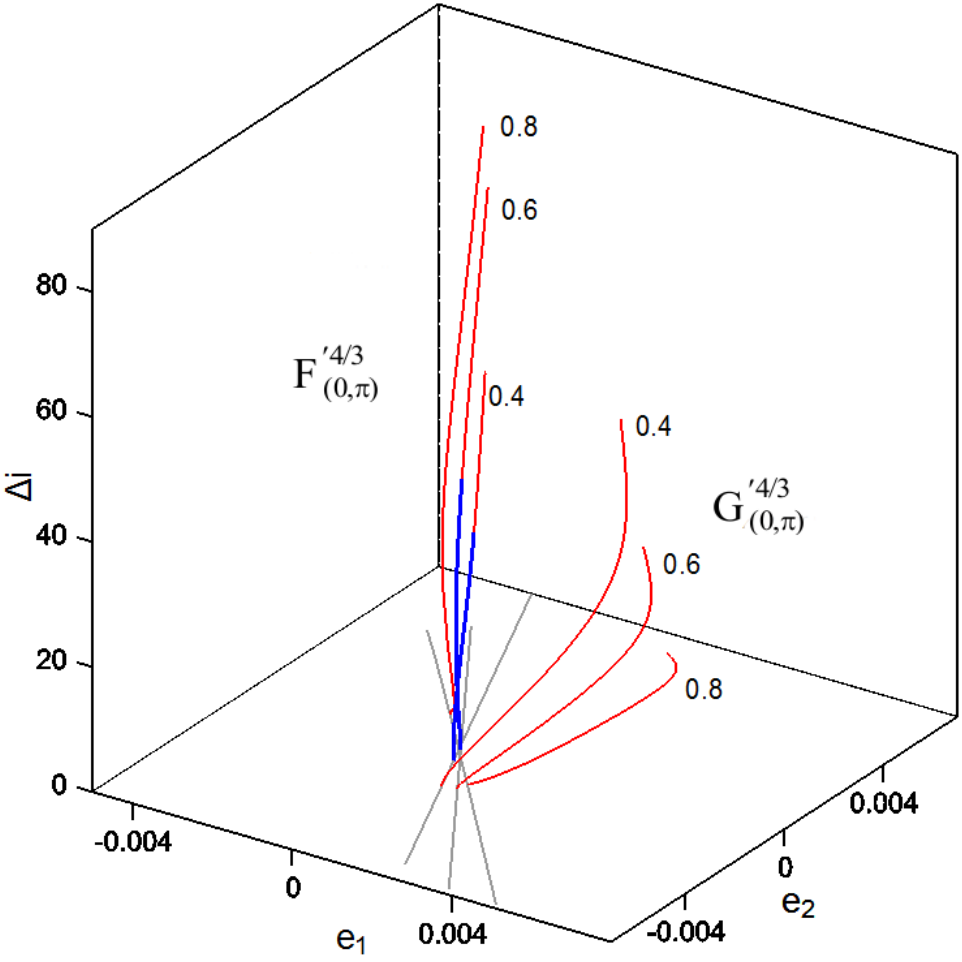} \vspace{-.4em} \\
\textnormal{(a)} & \textnormal{(b)} & \textnormal{(c)} \\ \vspace{-1.5em} 
\end{array} $
\end{center}\vspace{-.5em}
\caption{Planar and spatial families of symmetric periodic orbits in $4/3$ resonance.
}\vspace{-2em}
\label{43s1}
\end{figure}

\begin{figure}[H]
\begin{center}
$\begin{array}{@{\hspace{-.5em}}ccc}
\includegraphics[width=5.5cm,height=5.5cm]{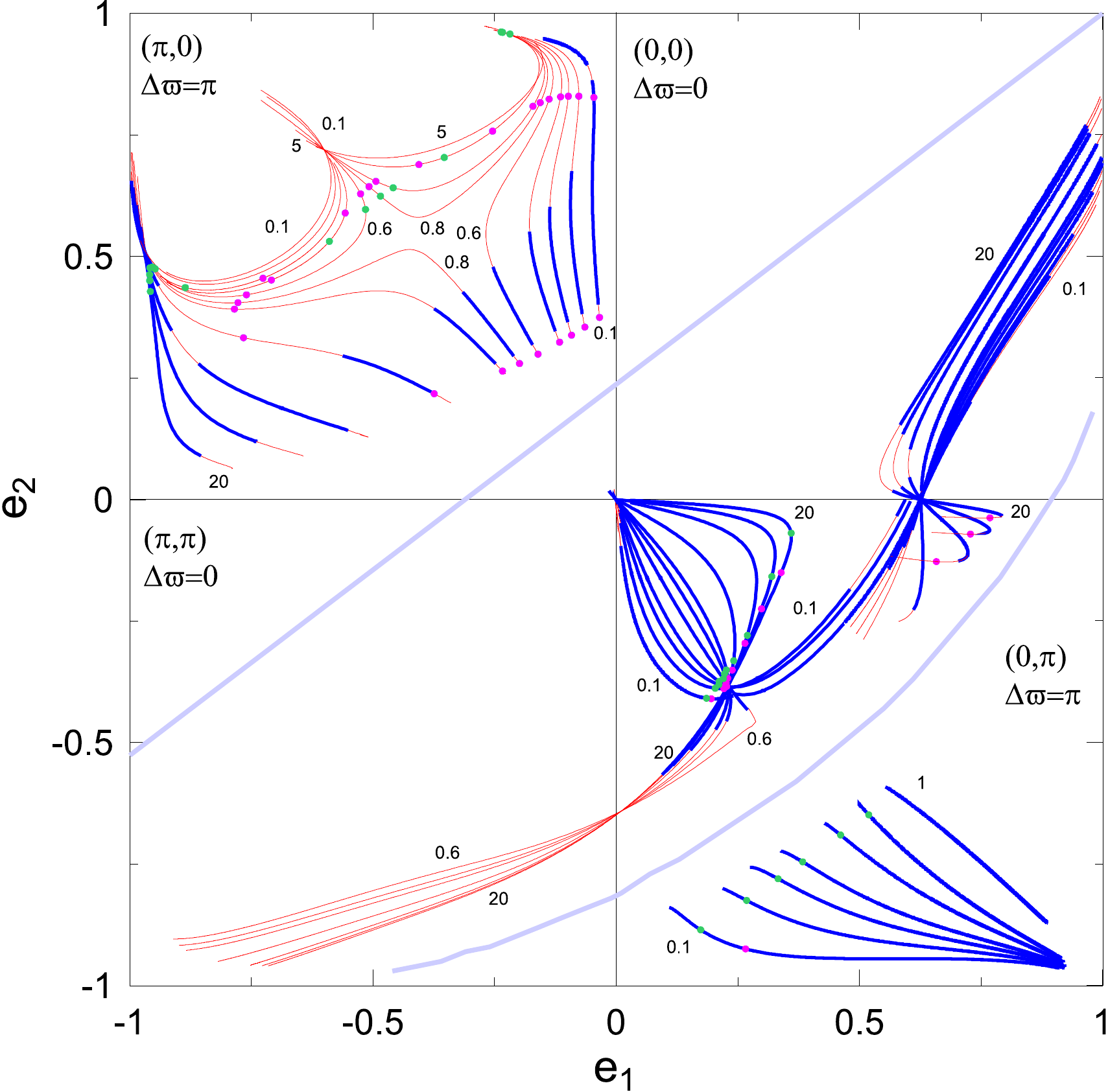}  &
\includegraphics[width=5.5cm,height=5.5cm]{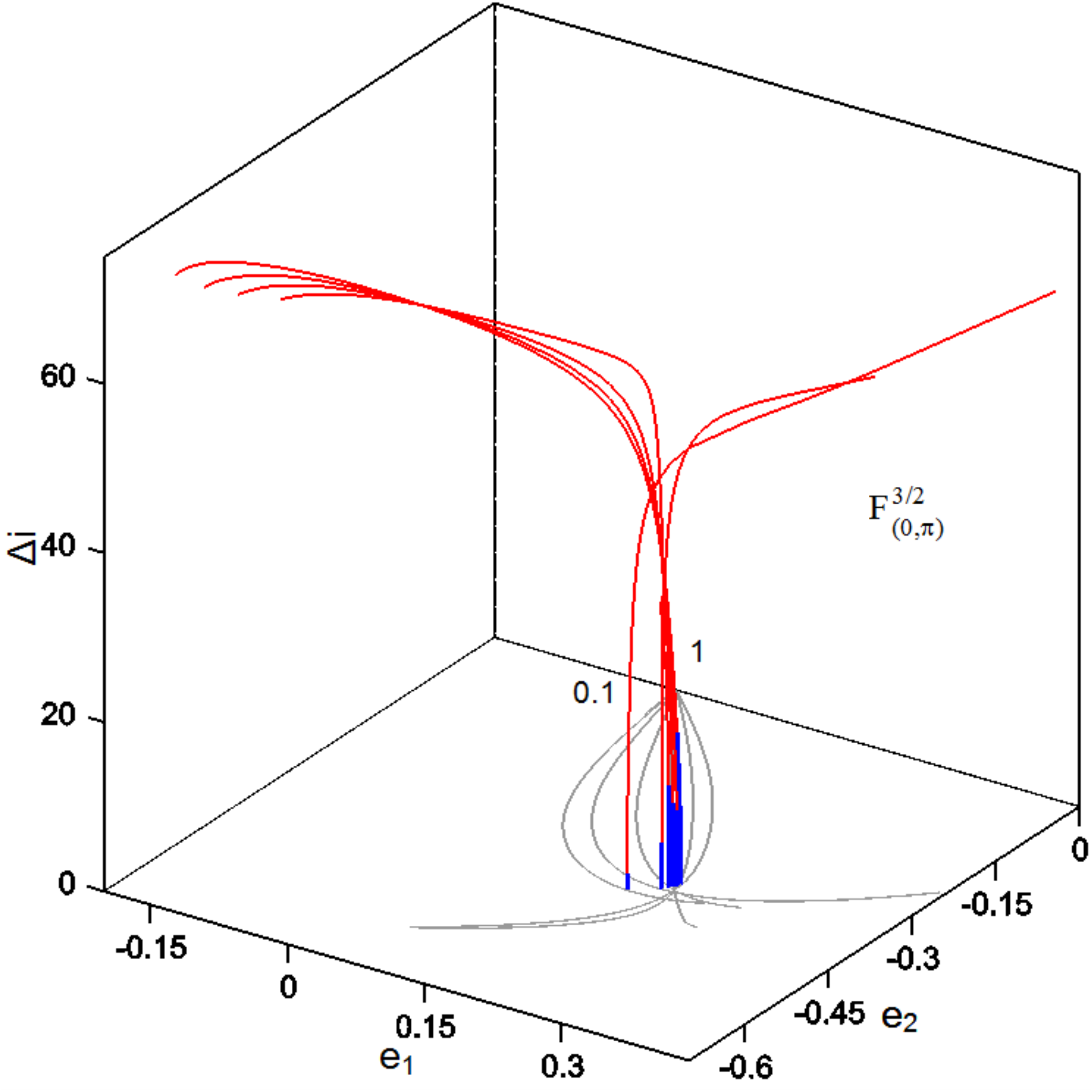}  &
\includegraphics[width=5.5cm,height=5.5cm]{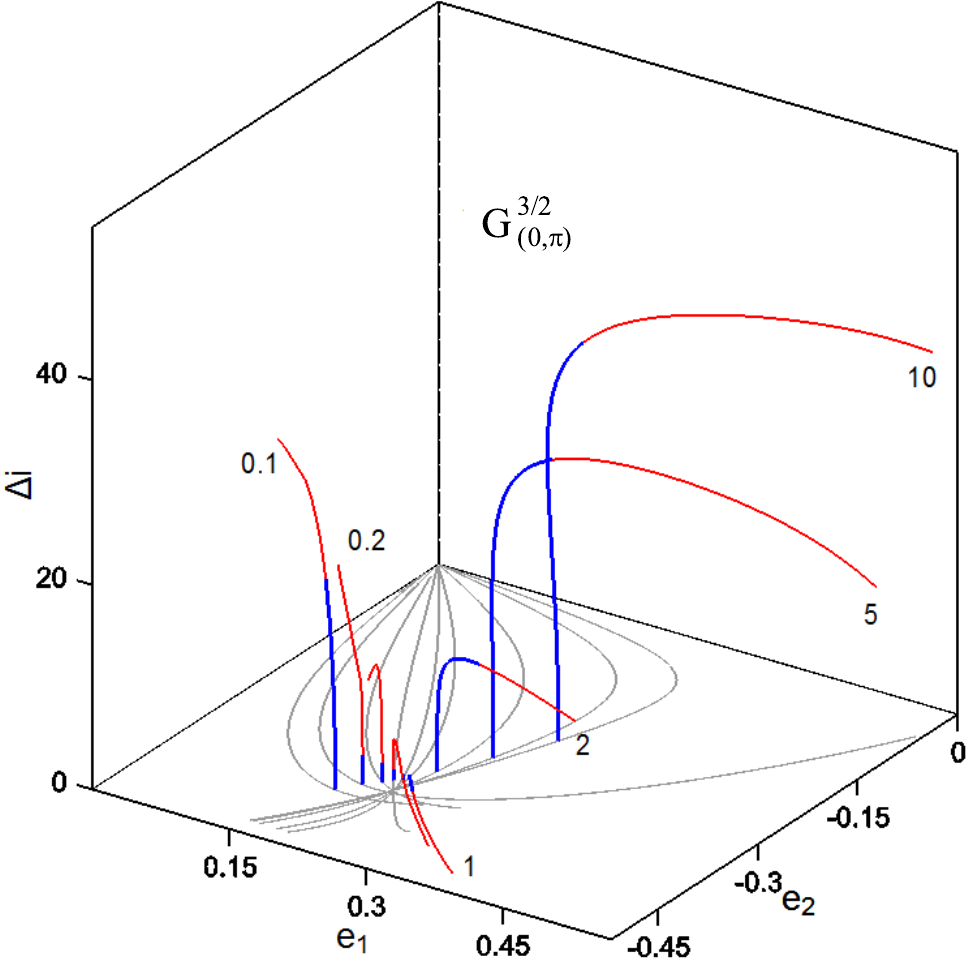} \vspace{-.4em}  \\
\textnormal{(a)} & \textnormal{(b)} & \textnormal{(c)} \\ \vspace{-1.5em} 
\end{array} $
\end{center}\vspace{-.5em}
\caption{Planar and spatial families of symmetric periodic orbits in $3/2$ resonance.}
\label{32s1}
\end{figure}

\vspace{-.5em} 
Following the literature \citep{mitch}, we project the families of planar periodic orbits in the eccentricity plane $(e_1,e_2)$. In order to distinguish the families of different configurations in the projection plane we use both positive and negative values for the eccentricities. The positive values of $e_i$ correspond to $\theta_i=0$ and the negative ones to $\theta_i=\pi$. We present the spatial families in the $3D$ projection space ($e_1,e_2,\Delta i)$, where $\Delta i$ is the mutual inclination of the planets given by the cosine rule
$\cos\Delta i=\cos i_1 \cos i_2+\sin i_1 \sin i_2\cos(\Omega_1-\Omega_2)$. We chose some examples of $4/3$ and $3/2$ MMRs (Figs. \ref{43s1},\ref{32s1}).
\vspace{-1.8em}

\vspace{-3em}

\section{Conclusions}\vspace{-.5em} 
We have performed an extensive study of $4/3$, $3/2$, $5/2$, $3/1$ and $4/1$ MMRs in the planar and spatial case of the GTBP, in an attempt to connect the dynamics of three dimensional periodic orbits with the evolution of multiplanetary exosystems found nowadays. The complete results of our study are given in \citep{av13}. Particularly, we provided the planar families of symmetric periodic orbits, which belong to all possible configurations that each MMR has, along with their linear horizontal and vertical stability. We computed the v.c.o. of these families and then, focused on continuing to space mainly the stable ones. We observe that both $xz$- and $x$-symmetric periodic orbits were found to be stable up to values of mutual inclination $50^\circ$-$60^\circ$, which could stimulate research of real systems whose planets are inclined. Also, we provided clues that could relate the required long-term stability of exoplanetary systems with three dimensional stable periodic orbits. In the neighbourhood of a stable periodic orbit, real planetary systems can be hosted and their long-term stability can be guaranteed. In contrary, if a planetary system is positioned in the vicinity of an unstable periodic orbit, due to the existence of chaotic regions around it, it will eventually destabilize. 

If families of periodic orbits can constitute paths that can drive the migration process of planets and finally, lock them in MMRs, this work can help determine and understand the reasons why, the exoplanets are discovered possessing certain attributes.

\noindent
{\bf Acknowledgements:} This research has been co-financed by the European Union (European Social Fund - ESF) and Greek national funds through the Operational Program "Education and Lifelong Learning" of the National Strategic Reference Framework (NSRF) - Research Funding Program: Thales. Investing in knowledge society through the European Social Fund.
\vspace{-1em}
\vspace{-.9em}
\end{document}